\documentclass[a4paper,11pt]{article}

\usepackage{jcappub} 

\usepackage[T1]{fontenc} 

\title{Decaying Dark Matter as a Possible Solution for Cosmological Tensions}

\author[a,b]{Yaman Acharya}
\author[a,1]{and Ryan E. Johnson\note{Corresponding author.}}

\affiliation[a]{Physics and Astronomy Department, Gettysburg College,\\300 N Washington St, Gettysburg, PA USA}
\affiliation[b]{University of California - Riverside,\\Riverside, CA, USA}

\emailAdd{yacha001@ucr.edu}
\emailAdd{rjohnson@gettysburg.edu}

\begin{document}
	
\abstract{
	Large-scale structure measurements have revealed persistent tensions between early- and late-time cosmological probes, most notably the long-standing discrepancy in the structure–growth parameter $S_8$. In this work, we explore how a model including decaying dark matter (DDM) can alleviate this tension by suppressing the growth of matter fluctuations at late times.  Specifically, we consider a neutrinophilic decay channel in which a heavy dark matter particle $\chi$ slowly decays into a Standard Model neutrino and a light invisible fermion, $\chi \rightarrow \nu + \phi$, modifying both the background evolution and the clustering of structure. Using the DES Year~1 redshift distributions, we construct a baseline matter power spectrum and compute the galaxy–galaxy, shear–shear, and galaxy–shear angular power spectra under both $\Lambda$CDM and DDM-inspired scenarios. We find that slow dark matter decay produces a scale-dependent suppression of clustering that remains consistent with DES measurements while naturally shifting the predicted structure amplitude toward the lower values favored by weak lensing surveys. Our results suggest that decaying dark matter is a compelling and physically motivated framework for addressing the $S_8$ tension. }

\maketitle
\flushbottom

\section{Introduction}
\subsection{$\Lambda$CDM and Cosmological Constraints}
\label{sec:intro}

Einstein introduced the cosmological constant $\Lambda$ in his field equations to allow for a static universe, but abandoned it after \textit{Edwin Hubble}’s 1929 discovery of cosmic expansion, later calling it his “biggest mistake.” Modern cosmology, however, reinterprets $\Lambda$ as a form of dark energy driving the observed accelerated expansion of the universe within the $\Lambda$CDM model, where dark energy and cold dark matter dominate cosmic dynamics and structure formation \citep{Davis_1985_evolution_large_scale_structure}. In parallel, the Big Bang theory provides the overarching framework for cosmic evolution, supported by evidence such as the cosmic microwave background, light element abundances, and large-scale galaxy distribution. It describes a universe that began in a hot, dense state and has been expanding and cooling for nearly 14 billion years, while also pointing to deeper questions about its unseen components.

\medskip

One of the most striking revelations of modern cosmology is that ordinary matter -- protons, neutrons, and electrons -- constitutes only about 5$\%$ of the total energy budget of the universe. Roughly 27$\%$ is attributed to dark matter, an invisible substance that does not interact with light but reveals itself through its gravitational influence on galaxies, clusters, and cosmic structure formation. The remaining 68$\%$ is thought to be dark energy, a mysterious component driving the accelerated expansion\citep{accel_de}.
The success of $\Lambda$CDM lies in its ability to incorporate these components and match a wide range of observations, from the anisotropies of the cosmic microwave background to the growth of cosmic structures. Nonetheless, the true nature of dark matter and dark energy remains among the greatest open questions in physics.

\medskip

In recent years, evidence has suggested that dark energy may not be a static cosmological constant but could evolve with time\citep{Copeland2006Dynamics}. Observations from large-scale surveys like Dark Energy Survey (DES) and the Dark Energy Spectroscopic Instrument (DESI) have hinted at departures from a perfectly constant $\Lambda$, pointing toward models of dynamical, or time-varying, dark energy. These evolving scenarios attempt to reconcile discrepancies in cosmological datasets and open the possibility of richer physics beyond $\Lambda$CDM however, this is not the only frontier. Tensions in cosmology, such as the well-known $H_0$ and $S_8$ tensions, highlight possible cracks in the standard $\Lambda$CDM model. \textit{The persistent $S_8$ tension in cosmology arises from a mismatch between the amplitude of matter clustering inferred from weak lensing surveys and that predicted by CMB observations}, motivating the exploration of new physics beyond $\Lambda$CDM. Although evolving dark energy presents a compelling path toward addressing current cosmological tensions, it cannot on its own account for all of the outstanding puzzles. Below, we elaborate on the $S_8$ tension, as it constitutes the central focus of this work. 

\medskip

One promising direction is the suite of evolving dark matter models, where dark matter itself may decay, interact, or transform in ways that affect structure growth across cosmic time. These possibilities suggest that the resolution to current cosmological puzzles may involve not only a dynamic dark energy component but also a deeper rethinking of the very nature of dark matter. A central motivation for exploring evolving dark matter models is the persistent discrepancy in the parameter ($S_8$), which measures the amplitude of matter clustering in the universe scaled by the growth of cosmic structures. The parameter is typically defined as
\begin{equation}
	S_8 = \sigma_8 \left( \frac{\Omega_m}{0.3} \right)^{0.5},
\end{equation}
where $\sigma_8$ quantifies the root-mean-square fluctuations of the matter density field on scales of $8 \, h^{-1} \, \mathrm{Mpc}$, and $\Omega_m$ is the present-day matter density parameter\citep{s8_tension}. It combines the matter clustering amplitude \(\sigma_8\) and matter density \(\Omega_m\) in a way that observational probes, such as weak lensing surveys, are most sensitive to. The matter fluctuations are captured by \(\sigma_8\) which measures the Root Mean Square (RMS) fluctuations on 8 \(h^{-1}\) Mpc scales, while \(\Omega_m\) affects the growth of structure. Surveys cannot independently constrain \(\sigma_8\) or \(\Omega_m\) well, but primarily constrain a degenerate combination, which is approximated by \(\sigma_8 (\Omega_m/0.3)^{0.5}\). The exponent 0.5 is chosen empirically to match this observational degeneracy\citep{s8_tension}

\medskip

\begin{figure}[tbp]
	\centering
	\hspace{-2.0cm} 
	\includegraphics[scale=0.38]{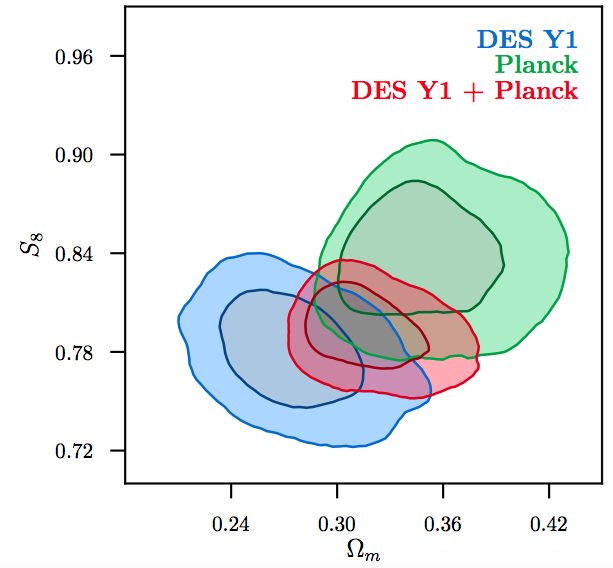}
	\caption{Comparison of $S_8$ from CMB and weak-lensing surveys, specifically DES Y1 data and Planck survey\citep{DES_tension}. The $S_8$ tension arising from mismatch between DES and CMBR measurements is represented, where the blue contours (DES Y1) and green contours (Planck) occupy statistically inconsistent regions in the $(\Omega_m, S_8)$ plane. Planck favors higher values of $S_8$ and a slightly larger matter density, while DES prefers lower amplitudes of structure, producing only partial overlap between the two. The combined DES+Planck constraints (red) fall in between but do not fully reconcile the discrepancy, illustrating the persistent nature of the $S_8$ tension in contemporary cosmology.}
	\label{fig:DESCRIPBABY}
\end{figure}

In the standard $\Lambda$CDM framework, predictions of $S_8$ are anchored by data from the cosmic microwave background (CMB), the relic radiation from the early universe\citep{planck_cos_18}. When we examine the late-time universe through probes of large-scale structure, however, a disagreement emerges. Weak gravitational lensing surveys, which trace how light from distant galaxies is bent by mass along the line of sight, tend to infer lower values of $S_8$ than those extrapolated from Planck’s CMB results under $\Lambda$CDM\citep{s8_tension}. This discrepancy, widely known as the "$S_8$ tension," signals either unaccounted-for systematics in data analysis or the need for new physics beyond the standard model of cosmology. The DES brought this tension to the forefront in its 2018 analysis, reporting $S_8$ values significantly below Planck predictions when combining galaxy clustering with weak lensing data\citep{DES2018} (see Figure\ref{fig:DESCRIPBABY}).

\medskip

More recent constraints in 2025, by both DES \citep{DES2025} and the DESI \citep{DESI2025}, have reinforced the tension with unprecedentedly precise measurements of the baryon acoustic oscillations and redshift-space distortions across millions of galaxies.  While both experiments' results improve constraints on cosmological parameters, they still leave a gap with Planck-calibrated $\Lambda$CDM predictions, underscoring that the $S_8$ tension is a persistent and robust feature across independent datasets.

\medskip

The DES and Planck measurements suggest that the discrepancy is not easily dismissed as statistical noise. Instead, they hint that late-time cosmic evolution may be governed by dynamics absent in the baseline $\Lambda$CDM picture. Evolving dark energy, interactions between dark matter and dark energy\citep{Boehmer2008}, or even models of decaying or self-interacting dark matter are increasingly being investigated as potential resolutions. The persistence of the $S_8$ tension thus stands as a compelling signal that alternative models may be worth investigating.

\medskip

In addition to these cosmological tensions, the formation of supermassive black holes in the early universe does not agree with the $\Lambda$CDM model of the universe. Observations reveal extremely massive quasars and supermassive black holes (SMBHs) at very high redshifts (e.g., \(z \gtrsim 6\)) that are difficult to reconcile with standard \(\Lambda\)CDM growth scenarios under Eddington-limited accretion. The usual timeline from seed black holes to the observed masses often does not leave enough cosmic time within the \(\Lambda\)CDM framework, especially given limits on accretion rates and gas supply\citep{bh_accret}. This, alongside cosmological tensions, remains a challenge that requires additional theoretical and observational insights.

\subsection{Dark Matter Models}

Though its true form remains elusive, we are able to place constraints on the nature of dark matter through its interactions with galaxies\citep{gal_dm}, clusters of galaxies\citep{clust_dm}, and the inter cluster medium\citep{icm_dm}.  
A particularly intriguing avenue of investigation is the possibility that dark matter is not entirely stable but instead decays into lighter particles. One class of models considers dark matter species that decay into a lighter daughter particle plus relativistic neutrinos. Such decays modify the evolution of the matter density across cosmic time: the abundance of clustering dark matter is reduced at late times, naturally lowering the amplitude of matter fluctuations and thereby addressing the long-standing $S_{8}$ discrepancy. For example, \citep{Pandey2019} demonstrate that allowing a fraction of dark matter to decay into lighter particles can simultaneously ease both the Hubble tension and the $\sigma_{8}$ anomaly, offering a unified framework for addressing several major cosmological challenges. The motivation for focusing on neutrino-producing decays lies in the resulting redistribution of energy density. While the stable fraction of dark matter continues to cluster, the relativistic neutrino products do not, instead free-streaming and suppressing the growth of structure. This brings late-time weak-lensing measurements into closer agreement with early-universe constraints such as those from the Planck CMB.

\medskip

In the broader landscape of Dark Matter decay scenarios, decay into neutrinos merits special attention over other decay channels since they naturally evade the gamma-ray and cosmic-ray bounds that exclude many visible-sector decay modes, yet still imprint measurable effects on the matter power spectrum through altered energy distribution. Such neutrino-producing decays often arise organically in models that link the dark sector to neutrino-mass generation, making them theoretically well-motivated as well as phenomenologically viable \citep{dm_neutrino_decay}.  
Moreover, neutrinos interact only weakly with ordinary matter, ensuring that such decay channels remain consistent with current observational bounds, making them a well-motivated candidate for addressing the $S_8$ tension.

\medskip

Analyses such as those by \citep{Olivares2017} investigate models in which dark matter interacts with neutrinos through scattering or decay channels, revealing that even small couplings can leave imprints on the growth of structures and the cosmic microwave background. More recent works, such as \citep{Mazoun2024}, extend this approach within the framework of interacting dark sectors, where ETHOS-inspired models allow for richer phenomenology connecting dark matter and neutrinos. These models suggest that the decay of higher-mass dark matter into lighter species plus neutrinos may account for the observed suppression of power on small scales without contradicting precision CMB constraints. Together, this body of work points to an emerging picture in which the apparent tensions in cosmology may not simply reflect systematic errors, but rather provide evidence of hidden dynamics in the dark sector.

\medskip

Further expanding on these ideas, \citep{FreeseWinkler2023} investigate a “Dark Big Bang” scenario, wherein a fraction of dark matter originates from a distinct, early injection of energy into the universe. Such scenarios naturally accommodate decaying or interacting dark matter, altering the evolution of density perturbations and potentially alleviating late-time cosmological tensions such as the $S_8$ discrepancy. Additionally, in \citep{FreeseWinkler2021}, the Chain Early Dark Energy model introduces sequential bursts of early dark energy that modify the expansion history of the universe, again highlighting the role of non-standard dark sector dynamics. Collectively, these works illustrate that decaying or interacting dark matter is a consistent and versatile framework capable of explaining multiple cosmological anomalies -- from early universe star formation to late-time structure growth -- while remaining compatible with current observational constraints. To connect these theoretical models with observations, one must turn to statistical probes of large-scale structure that encode how matter clusters across cosmic time.

\subsection{Matter Power Spectrum}
A key statistical tool for characterizing the distribution of matter in the universe across different spatial scales is the matter power spectrum, \(P(k)\).  By analyzing fluctuations in the density of galaxies, clusters, and the underlying dark matter, \(P(k)\) allows us to quantify how structures grow over cosmic time and to test cosmological models. On the theoretical front, \(P(k)\) encodes the variance of the matter density contrast \(\delta(\mathbf{x})\) in Fourier space, capturing both the amplitude and scale dependence of density perturbations.  At the same time we may use observations from galaxy surveys, weak gravitational lensing, and the cosmic microwave background to provide empirical constraints to \(P(k)\) and its predictions from $\Lambda$CDM or alternative cosmologies. These comparisons reveal subtle discrepancies, such as the suppression of clustering at certain scales, which can indicate new physics in the dark sector, including scenarios where dark matter decays or interacts with relativistic species. By studying the power spectrum in detail, we are able to constrain parameters like the matter density \(\Omega_m\), the amplitude of fluctuations \(\sigma_8\), and the properties of dark matter and dark energy, making it an indispensable tool for understanding the large-scale structure and evolution of the universe.

\medskip

In particular, we use the \textbf{power spectrum of lens galaxies} ($C_{\ell}^{gg}$), \textbf{source galaxies} ($C_{\ell}^{\kappa\kappa}$), and \textbf{lens--source correlations} ($C_{\ell}^{\kappa g}$) to study the large-scale structure of the Universe\citep{DES2018}. \textbf{Lens galaxies} are foreground galaxies whose spatial distribution traces the underlying matter density field, and $C_\ell^{gg}$ quantifies the clustering of these galaxies as a function of angular scale. \textbf{Source galaxies} are more distant background galaxies whose observed shapes are distorted by gravitational lensing from matter along the line of sight. The angular power spectrum $C_\ell^{\kappa\kappa}$ measures these weak lensing distortions (cosmic shear) across the sky. The cross-correlation between lens and source galaxies, $C_\ell^{\kappa g}$, captures \textbf{galaxy--galaxy lensing}, probing how the mass associated with lens galaxies distorts the shapes of background sources, connecting galaxy clustering and gravitational lensing. Together, these observables encode crucial information about the distribution of matter, the processes of galaxy formation and evolution, and the underlying cosmological parameters that govern the expansion of the Universe.  

\medskip

These calculations often employ the \textbf{Limber approximation}, which assumes that the weighting (or kernel) functions vary slowly over the relevant scales so that the oscillatory Bessel integrals can be approximated by delta functions \citep{limber1953analysis}. This simplifies the projection of the three-dimensional matter power spectrum to the two-dimensional angular power spectra measured on the sky. Under this approximation, the angular power spectrum between two tracers, $X$ and $Y$, is given by  

\begin{equation}
	C_{\ell}^{XY} = \int_{0}^{\infty}\frac{dz}{c}\frac{H(z)}{\chi^2(z)}[W^{X}(z) W^{Y}(z)] P\bigg(k=\frac{\ell}{\chi(z)},z \bigg),
	\label{eq:powerspe_th}
\end{equation}
where $\chi(z)$ is the comoving distance to redshift $z$, $H(z)$ is the Hubble parameter, and $P(k,z)$ is the three-dimensional matter power spectrum. Physically, this equation describes how matter density fluctuations at different redshifts project onto angular scales on the sky, weighted by the sensitivity (or kernel) functions of the tracers involved.  

The function $W(z)$ denotes the \textbf{redshift-dependent kernel} of each tracer \citep{marques2024cosmological}. For \textbf{lens galaxies}, assuming a linear and deterministic galaxy bias, the kernel is  

\begin{equation}
	W^{g}(z) = b(z) \frac{dn}{dz},
\end{equation}
where $b(z)$ represents how galaxies trace the underlying matter distribution, and $dn/dz$ is the normalized redshift distribution of the lens population. This kernel enters directly in the calculation of $C_\ell^{gg}$ and $C_\ell^{\kappa g}$, linking galaxy clustering to the matter density.

\medskip

For \textbf{source galaxies}, the \textbf{weak lensing kernel} $W^{\kappa}(z)$ characterizes how the foreground matter distribution distorts the shapes of background galaxies through gravitational lensing:  

\begin{equation}
	W^{\kappa}(z) =  \frac{3 H^2_0 \Omega_{\rm{m},0}}{2 H(z) c} \frac{\chi(z)}{a(z)} \int^{z_H}_{z} dz' n(z') \frac{\chi(z') - \chi(z)}{\chi(z')},
\end{equation}

where $n(z)$ is the redshift distribution of source galaxies, $\Omega_{\rm m,0}$ is the present-day matter density, $H_0$ is the Hubble constant, and $a(z)$ is the scale factor. This kernel effectively weights how efficiently \textbf{structures at redshift $z$ lens background sources at redshift $z'$}, linking the geometry and growth of structure in the Universe. It enters directly into $C_\ell^{\kappa\kappa}$, describing the cosmic shear signal, and into $C_\ell^{\kappa g}$, quantifying the lensing by foreground galaxies\citep{marques2024cosmological}. 

\medskip

\subsection{Cosmological Implications from Angular Power Spectra}

The angular power spectra computed from the DES Y1 tomographic bins serve as a direct probe of the underlying matter distribution and the influence of dark energy on the growth of cosmic structure. In particular, the galaxy clustering ($C_\ell^{gg}$) and weak lensing ($C_\ell^{\kappa\kappa}$) spectra encode complementary information about the distribution of dark matter, while the galaxy–galaxy lensing cross spectrum ($C_\ell^{\kappa g}$) links the positions of galaxies to the surrounding matter density field.

\medskip

Furthermore, the weak lensing and galaxy-galaxy lensing spectra provide a direct link between the luminous tracers (galaxies) and the underlying dark matter distribution. By comparing the auto- and cross-correlations across tomographic bins, we are able to investigate how matter fluctuations evolve with redshift and how dark energy affects the geometry of the Universe. This methodology allows the study of changes in the large-scale structure of the cosmos, including the growth rate of dark matter halos, the scale dependence of clustering, and the correlation between matter and luminous tracers.

\medskip

By systematically analyzing these power spectra under different cosmological models, we are able to place constraints on the nature of dark matter, assess the impact of evolving dark matter on cosmic structure formation, and ultimately connect the observational signatures of galaxy clustering and weak lensing to fundamental properties of the Universe.

\section{Method}

\subsection{Galaxy Distributions and Redshifts}

Our analysis uses publicly available data from the \textbf{Dark Energy Survey (DES) Year 1 (Y1)} release to construct and visualize the redshift distributions of the lens and source galaxy samples.  DES is a wide-field optical survey that has mapped approximately 5000 deg$^2$ of the southern sky in five photometric bands ($grizY$). The survey provides precise photometric redshifts for millions of galaxies, enabling measurements of galaxy clustering, weak gravitational lensing, and other cosmological probes. We specifically used the Y1 redshift distributions for both lens and source galaxy samples as described in Ref.\citep{DES_Y1_galgal}. By working directly with the DES Y1 tomographic binning and redshift distributions, we ensured that our analysis is consistent with observationally derived galaxy samples and realistic survey conditions.

\medskip

\begin{figure}[tbp]
	\centering
	\includegraphics[scale=0.7]{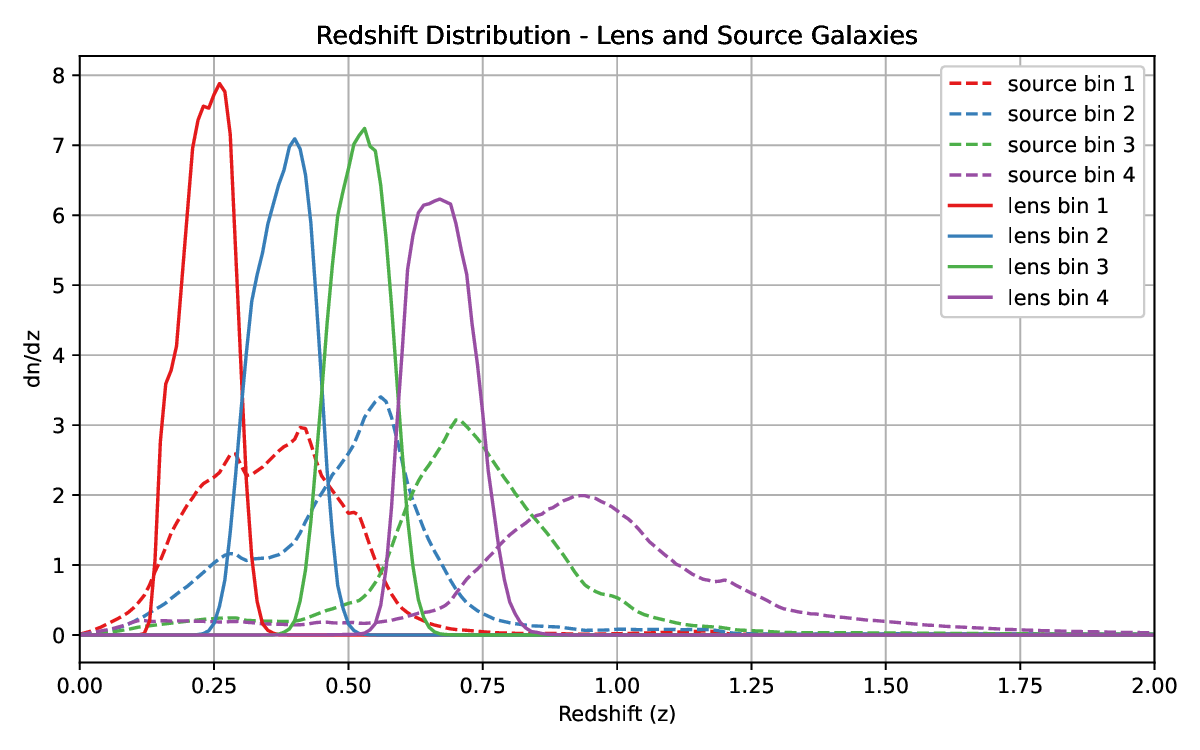}
	\caption{Redshift distributions ($\frac{dN}{dz}$) of the lens (solid lines) and source (dashed lines) galaxies from the DES Year 1 dataset. Each curve represents one tomographic bin, showing how the galaxy samples are distributed as a function of redshift.}
	\label{fig:dndz}
\end{figure}

The \textit{lens galaxies} correspond to those used for galaxy clustering measurements, while the \textit{source galaxies} are those used for weak lensing (cosmic shear) studies. Both samples are divided into \textbf{tomographic redshift bins}, which are intervals in redshift space that group galaxies according to their estimated distances. This tomographic binning allows cosmological information to be extracted as a function of redshift, helping to trace the growth of structure and the geometry of the Universe over cosmic time.  Figure~\ref{fig:dndz} shows the redshift distributions of both the source and lens galaxies.

To compute theoretical predictions for these observables, we utilized the \texttt{pyccl} library, the Python interface to the Core Cosmology Library (CCL)\citep{Chisari2018_CCL}$^,$\citep{CCL_code}. This library provides a consistent and efficient framework for modeling large-scale structure observables within a given cosmological model. For an input set of cosmological parameters, such as $\Omega_m$, $\sigma_8$, and the dark energy equation of state, \texttt{pyccl} computes the matter power spectrum $P(k,z)$ using established prescriptions for both linear and nonlinear structure formation. It then projects this three-dimensional information into angular power spectra $C_\ell$ through line-of-sight integrations, enabling direct comparison with observational data. The tomographic binning was implemented through redshift-dependent window functions, ensuring consistency with the survey’s observational setup. Additionally, the framework allows for the inclusion of galaxy bias and nonlinear corrections, which are essential for accurately modeling structure growth on relevant scales. This approach enables the generation of realistic theoretical predictions that can be directly compared with DES measurements.

\medskip

\subsection{Baseline Power Spectrum}
\label{base}

To establish a reference model, we first construct the linear matter power spectrum using the DES Year~1 best-fit cosmological parameters. We fixed the matter and baryon densities, Hubble parameter, and other standard cosmological parameters to representative DES values, which are specified below. 

\[
\begin{aligned}
	\Omega_c &= 0.226, \\
	\Omega_b &= 0.0493, \\
	\Omega_m &= \Omega_c + \Omega_b, \\
	h       &= 0.7, \\
	A_s     &= 2.1 \times 10^{-9}, \\
	n_s     &= 0.96, \\
	\Omega_k &= 0.0, \\
	w_0     &= -1.0, \\
	w_a     &= 0.0.
\end{aligned}
\]

These parameters are implemented in \texttt{pyccl} as our baseline cosmology. This model reproduces the matter distribution consistent with DES Y1 analyses and serves as the benchmark against which all subsequent modifications are compared (see black lines in Fig.\ref{fig:wababy}).  The resulting linear matter power spectrum captures the key clustering properties of the standard $\Lambda$CDM model and forms the foundation of our decaying dark matter analysis.

\medskip

The angular power spectra were computed over logarithmically spaced multipoles using the \texttt{ccl.angular\_cl} routine, and plotted on logarithmic scales to highlight the variation across large and small angular scales. These allow investigation of how the matter and lensing power varies across different angular scales and redshift bins, providing a basis for comparison with theoretical models and future survey predictions.

\medskip

Using the DES Y1 redshift distributions for both lens and source galaxies, we compute the angular power spectra for each tomographic bin. We first extract the $\frac{dN}{dz}$ distributions from the DES Y1 FITS release for all four tomographic bins of both source and lens galaxies. The source galaxy distributions were used for weak lensing tracers, while the lens galaxy distributions were used for number count tracers with an assigned galaxy bias, $b(z)$, for each bin.

\medskip

In our subsequent analysis, we employed the \textbf{Core Cosmology Library (CCL)}\citep{Chisari2018_CCL} in python and compute the following angular power spectra:

\begin{enumerate}
	\item \textit{Weak Lensing Auto-Power Spectrum} $C_\ell^{\kappa\kappa}$: Using the source galaxy $\frac{dN}{dz}$ for each tomographic bin, we compute the auto-correlation of the weak lensing convergence field across angular scales $\ell = 2$--$2000$. These spectra quantify the projected cosmic shear signal and its evolution with redshift.
	
	\item \textit{Galaxy Clustering Auto-Power Spectrum} $C_\ell^{gg}$: Using the lens galaxy $\frac{dN}{dz}$ and the corresponding galaxy bias for each bin, we compute the auto-correlation of galaxy number counts. This spectrum characterizes the clustering of galaxies within each tomographic bin.
	
	\item \textit{Galaxy–Galaxy Lensing Cross Power Spectrum} $C_\ell^{\kappa g}$: Finally, we compute the cross-correlation between lens galaxies and source galaxy weak lensing tracers. This measurement probes the matter distribution around galaxies and the lensing signal induced by foreground structures.
\end{enumerate}

\subsection{Particle-physics framework for decaying dark matter}
\label{subsec:theory_DM_decay}

As a starting point, we adopt a neutrinophilic decaying DM scenario in which a heavy DM particle $X$ with
mass $m_X \sim \mathcal{O}(\mathrm{PeV})$ is metastable on cosmological timescales and decays predominantly into Standard Model (SM) neutrinos and
invisible dark-sector states. This setup follows the general framework of \citep{Hiroshima2018}, where multi-body DM decays are used to explain the diffuse high-energy neutrino flux observed by IceCube while satisfying multi-messenger constraints.

\medskip

We assume that $X$ is a SM gauge singlet (scalar or fermion) whose dominant interactions with the visible sector are described at low energy by effective operators that couple $X$ to SM lepton doublets and to a set of dark-sector fermions $\psi_i$. At the renormalizable level, a simple two-body decay channel can be written schematically as

\begin{equation}
	\mathcal{L}_{\mathrm{int}}^{(2)}
	\;=\;
	y_\alpha \, X \, \bar{\nu}_\alpha \psi
	\;+\; \mathrm{h.c.} \,,
	\label{eq:Lagrangian_two_body}
\end{equation}

where $\nu_\alpha$ ($\alpha = e,\mu,\tau$) denotes the SM neutrino fields, $\psi$ is a neutral dark-sector fermion, and $y_\alpha$ are small Yukawa couplings. The corresponding partial decay width for the two-body channel $X \rightarrow \nu_\alpha + \psi$ is

\begin{equation}
	\Gamma_{2}
	\;\simeq\;
	\frac{|y_\alpha|^2}{16\pi} \, m_X
	\left(1 - \frac{m_\psi^2}{m_X^2}\right)^2 ,
	\label{eq:Gamma_two_body}
\end{equation}

where we have neglected neutrino masses. The requirement that $X$ constitutes the dark matter today implies a lifetime $\tau_X = 1/\Gamma_{\mathrm{tot}} \gg t_0$, where $t_0$ is the age of the Universe. In practice, the neutrino data considered in Ref.\citep{Hiroshima2018} favor lifetimes in the range $\tau_X \sim 10^{26}–10^{28}\,\mathrm{s}$ for $m_X \sim \mathrm{PeV}$, corresponding to very small couplings $|y_\alpha|$.

\medskip

To obtain a broader neutrino energy spectrum, we also allow for a multibody decay channel in which the DM particle decays into one neutrino and
$N-1$ invisible dark-sector fermions,

\begin{equation}
	X \rightarrow \nu_\alpha + \psi_1 + \psi_2 + \cdots + \psi_{N-1} \,.
\end{equation}

Such decays can arise from higher-dimensional operators suppressed by a large mass scale $\Lambda$,

\begin{equation}
	\mathcal{L}_{\mathrm{int}}^{(N)}
	\;\sim\;
	\frac{1}{\Lambda^{\,n-4}}\,
	X\,\nu_\alpha \,\psi_1 \psi_2 \cdots \psi_{N-1}
	\;+\; \mathrm{h.c.} \,,
	\label{eq:Lagrangian_multi_body}
\end{equation}

where $n$ is the operator dimension. After integrating over the $N$-body phase space in the rest frame of $X$, this interaction produces a continuous neutrino spectrum $dN_\nu/dE_\nu$ extending from $E_\nu \simeq 0$ up to $E_\nu \simeq m_X/2$. We may treat this spectrum using the parameterizations provided in Ref.\citep{Hiroshima2018}, where the shape is controlled by the effective multiplicity $N$ and the relative branching fractions of the two-body and multibody channels.

\medskip

The total decay width of $X$ is then written as

\begin{equation}
	\Gamma_{\mathrm{tot}}
	\;=\;
	\Gamma_{2} + \Gamma_{N} \,,
\end{equation}

and we define the corresponding branching ratios $\mathrm{BR}_{2} = \Gamma_{2}/\Gamma_{\mathrm{tot}}$ and $\mathrm{BR}_{N} = \Gamma_{N}/\Gamma_{\mathrm{tot}}$. In the limit where the final state consists only of neutrinos and dark-sector fermions, electromagnetic and hadronic secondaries are strongly suppressed, so the model can evade stringent $\gamma$-ray constraints on DM-induced neutrino production (see the detailed discussion in Ref.~\cite{Hiroshima2018}).

\section{Results and Discussion}
\label{sec:results}

\subsection{Sensitivity of Large-Scale Structure (LSS) Power Spectra to the Equation-of-State Parameter \(w_a\)}

\begin{figure}[tbp]
	\noindent
	\hspace{-1.3cm} 
	\includegraphics[scale=0.57]{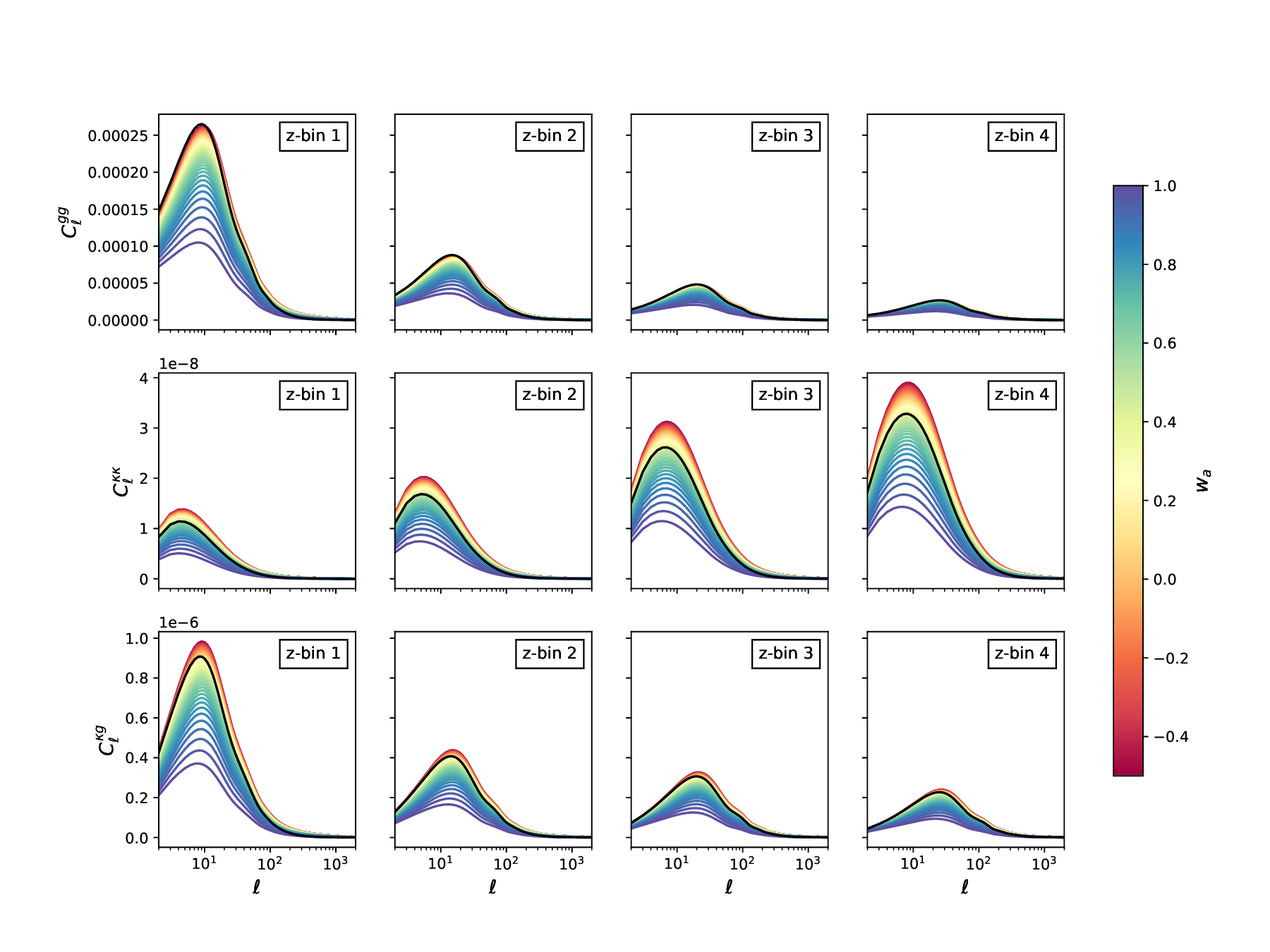}
	\vspace{-35pt} 
	\caption{In this figure we show the auto- and cross-correlated matter power spectra from the DES redshift sample.  We vary the dark energy evolution parameter $w_a$, demonstrating the sensitivity of cosmic structure growth to changes in cosmological parameters. The black line in each subplot shows our baseline matter power spectrum constructed using DES Y1 cosmological parameters, serving as the reference model against which all modified scenarios are compared.}
	\label{fig:wababy}
\end{figure}

We first explore how the power spectra from LSS observables are sensitive to changes in the dark energy equation of state parameter \(w_a\). Keeping all other cosmological parameters fixed, we vary \(w_a\) and compute the resulting matter power spectra.  As shown in Fig.~\ref{fig:wababy}, even modest changes in \(w_a\) produce noticeable deviations in the power spectrum on intermediate and large spatial scales (intermediate and small $\ell$-values). This highlights the strong response of structure growth to changes in cosmological parameters, illustrating that any physical mechanism that alters the background or perturbation evolution -- including DDM -- can leave detectable imprints on $P(k)$.  It is this sensitivity that motivates our use of power-spectrum-based statistics to probe dark matter decay.

\medskip

\subsection{Why the Cross-Correlation \texorpdfstring{$C_\ell^{\kappa g}$}{Clkg} is the Key LSS Observable}

Although both the galaxy auto-spectrum $C_\ell^{gg}$ and the lensing auto-spectrum $C_\ell^{\kappa\kappa}$ contain information about structure formation, they each suffer from significant degeneracies or observational limitations:

\begin{itemize}
	\item $C_\ell^{gg}$ is strongly affected by galaxy bias.
	\item $C_\ell^{\kappa\kappa}$ is limited by observational noise in weak-lensing 
	reconstruction.
\end{itemize}

The galaxy--lensing cross-correlation $C_\ell^{\kappa g}$ mitigates both issues by linking the biased galaxy field with the unbiased underlying matter field. This breaks degeneracies in galaxy bias and provides a more direct probe of matter clustering. Consequently, $C_\ell^{\kappa g}$ is the most reliable observable for isolating the impact of DDM on large-scale structure, and we adopt it as the primary statistic for testing deviations from $\Lambda$CDM.

\subsection{Coupling Baryons and Dark Matter at Fixed Total Density}

To investigate how DDM-like scenarios affect clustering while preserving the large-scale structure of the Universe, we vary the \textit{relative contributions} of baryons and cold dark matter while keeping the total matter density fixed:

\[
\Omega_m = \Omega_b + \Omega_c = 0.3.
\]

By allowing $\Omega_b$ to increase and $\Omega_c$ to decrease with time, this serves as our simplified model of partial DM decay into baryons, representing a slow transfer of mass between components without altering the overall matter density.

\medskip

\begin{figure}[tbp]
	\noindent
	\hspace{-1.5cm} 
	\includegraphics[scale=0.57]{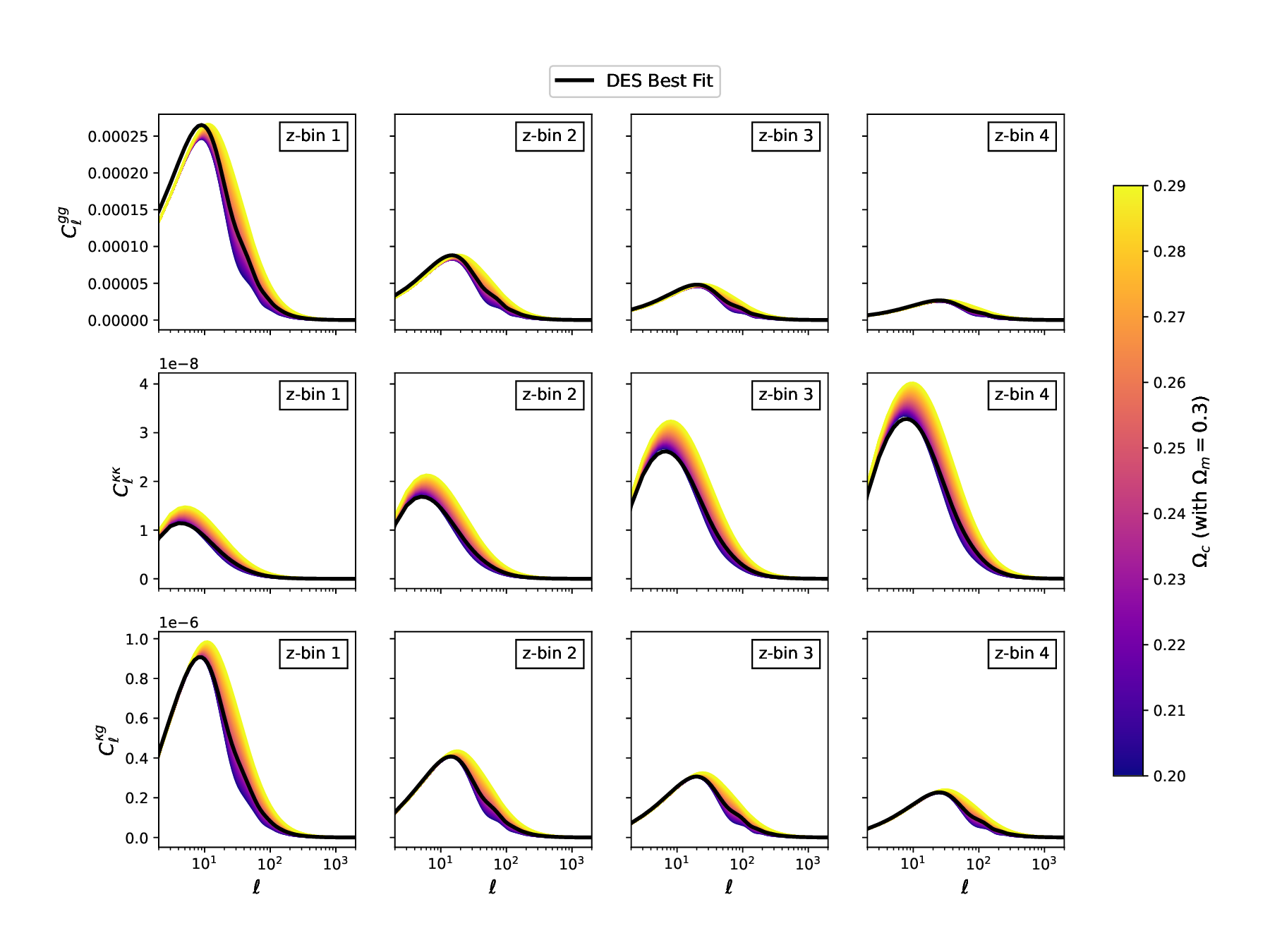}
	\caption{An illustration of the effect of our decaying dark matter model on the resulting auto- and cross-correlated matter power spectra.  This shows the impact that redistributing matter between dark matter and baryons, while keeping $\Omega_m$ fixed, has on the matter power spectrum. Large-scale structure remains nearly unchanged, with only mild small-scale suppression.}
	\label{fig:coupling_figure}
\end{figure}

Figure~\ref{fig:coupling_figure} shows that such variations lead to minimal changes in the matter power spectrum on large scales (small $\ell$), preserving the overall shape of structure formation. Only mid-scale power changes noticeably due to the reduced dark matter fraction. This behavior mirrors the expected impact of slow DDM, which weakens clustering but does not significantly alter the largest-scale features.

\medskip

If the tomographic binning permits a time evolution discussion, we note that impact of varying $\Omega_c$ appears most significant on the redshift bin 1 data, and that effect lessens with increasing redshift.

\subsection{Comparison with DES Cross-Correlation Measurements}

\begin{figure}[tbp]
	\noindent
	\hspace{-1.4cm} 
	\includegraphics[scale=0.57]{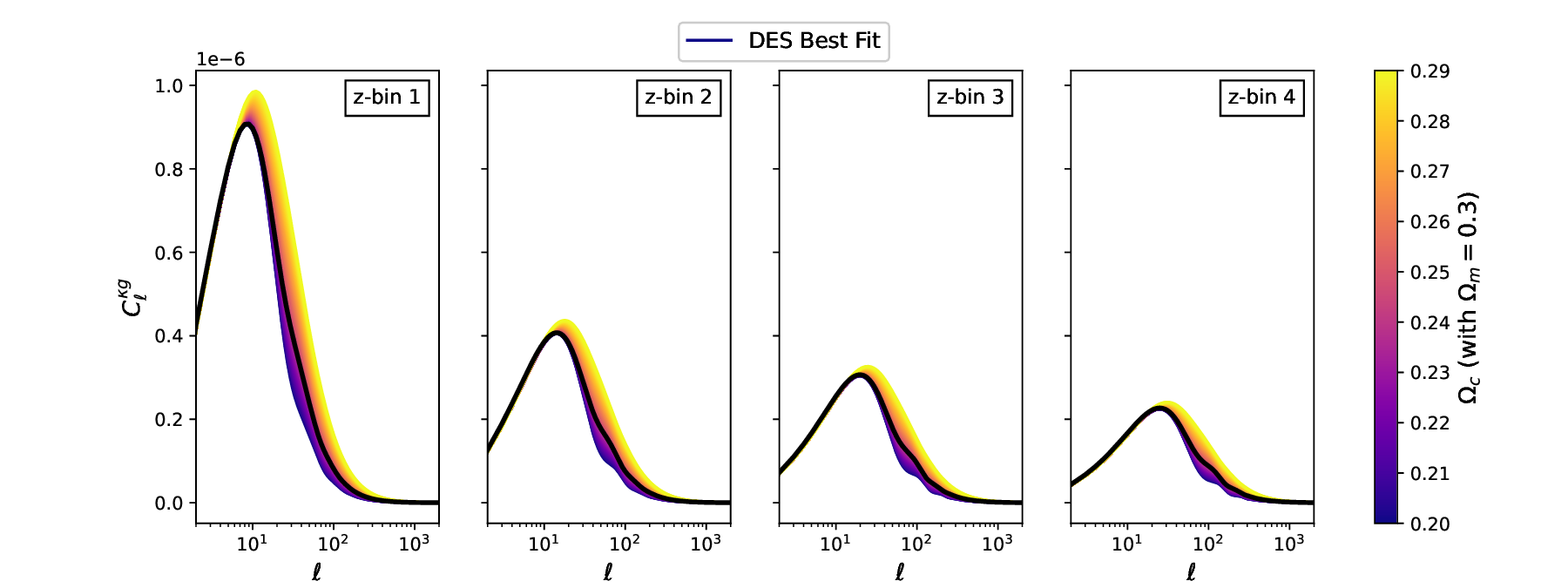}
	\caption{Comparison of the DES Y1 best-fit galaxy–lensing cross-correlation $C_\ell^{\kappa g}$ with models in which dark matter and baryons are rebalanced at fixed $\Omega_m$. The agreement on large scales and the small deviations on smaller scales illustrate that such scenarios can preserve the overall structure while mildly suppressing clustering.}
	\label{fig:DES_comparision}
\end{figure}

Finally, we compare our theoretical predictions for $C_\ell^{\kappa g}$ (the galaxy-lensing cross-correlation coefficient) with DES Year 1 measurements. In the context of decaying dark matter, this observable becomes especially informative: since the decay primarily affects the growth of structure at late times (zbin 1), it preserves the amplitude of large-scale modes while naturally suppressing the small-scale power where the decay-induced velocity kicks and energy redistribution become important. As shown in Fig. \ref{fig:DES_comparision}, the resulting scale-dependent suppression produces a characteristic downturn in $C_\ell^{\kappa g}$ at high multipoles, reflecting the reduced clustering efficiency of the dark matter component while maintaining consistency with observations on larger angular scales.

\medskip

The agreement on large scales reinforces our earlier result: transferring mass between dark matter and baryons -- or equivalently, cosmological scenarios in which dark matter slowly decays -- preserves the key features of large-scale structure. On smaller scales ($\ell \gtrsim 400$), the model predicts only mild departures from the DES best-fit curve, consistent with the expectation that DDM produces a small suppression of clustering. These deviations should remain within observational uncertainties, implying that current DES measurements cannot rule out such DDM-inspired scenarios, although more data on error is required to confirm this, and future surveys may significantly tighten constraints.

\section{Conclusion and Future Work}

We present in this work a preliminary investigation into the role of decaying dark matter (DDM) as a potential resolution to the long-standing $S_{8}$ tension in contemporary cosmology. By modeling the impact of dark matter decay on the growth of large-scale structure and comparing the resulting predictions with the galaxy--lensing cross-correlations from the DES Y1 data set, we find that DDM-induced suppression of clustering offers a physically motivated and quantitatively plausible mechanism for reconciling discrepancies between weak-lensing and CMB-derived structure amplitudes. Although the current results are promising, they remain exploratory in nature. 

\medskip

This framework allows for a direct calculation of $S_{8}$ from the DDM-modified matter power spectrum, which will be the focus of subsequent work.  In practice, determining whether the decay can fully remove the $S_{8}$ tension requires a full Bayesian exploration of parameter space using MCMC methods\citep{Akeret2013_CosmoHammer}. The key model parameters controlling the suppression of late-time structure -- the dark-matter lifetime $\tau_{X}$, the branching fractions of the decay channels, and the effective multiplicity $N$ governing the neutrino spectrum -- can be tuned within observational limits to reduce the small-scale power in a way compatible with DES measurements. Preliminary power-spectrum results already indicate that DDM naturally weakens late-time clustering, suggesting that with a full statistical inference pipeline, the model may indeed yield an $S_{8}$ prediction consistent with low-redshift probes. 

\medskip

We also plan to extend our analysis to include the cross-correlation between DES galaxy clustering and IceCube high-energy neutrino data. Identifying a shared spatial structure between these two tracers has the potential to provide an independent probe of dark matter decay, as neutrinos produced in such scenarios should trace the underlying matter distribution. In terms of observational constraints, there is room to expand our analysis by incorporating other wide-area surveys with significantly larger sky coverage, such as LSST, \textit{Euclid}, and future CMB lensing maps. These next-generation data sets offer improved sensitivity, reduced noise, and more homogeneous coverage, all of which are essential for disentangling genuine physical correlations from statistical fluctuations or residual systematics.

\medskip

Although substantial work remains, the preliminary results of our exploration are promising.  Should our forthcoming analyses confirm the presence of correlated structure between galaxy surveys and neutrino observatories consistent with DDM predictions, it would have far-reaching implications. Such a result would not only help to address one of the most persistent tensions in modern cosmology but would also provide unprecedented insight into the nature of the dark sector of the universe. Establishing evidence for dark matter decay would fundamentally reshape our understanding of cosmic evolution, guide the development of next-generation dark matter detection strategies, and impose powerful new constraints on particle physics beyond the Standard Model.

\medskip

\begin{acknowledgments}
We gratefully acknowledge the \textsc{DES Collaboration} for providing access to the public Year~1 data products used in this analysis.  This work also benefited from computational tools, discussions, and resources made possible through the support of \textsc{Fermilab} and its Department of Particle Astrophysics.  We also thank Gettysburg College's \textsc{Kolbe} and XSIG research funds, without which this work could not have progressed to its current state.  
	
\end{acknowledgments}

\end{document}